Impact of EEG biofeedback on event-related potentials (ERPs) in attention-deficit hyperactivity (ADHD) children.

S. Bakhtadze1, M. Janelidze1, N. Khachapuridze2.

1. Department of clinical neurology, Tbilisi State Medical University, Tbilisi, Georgia.

2. Department of pediatric neurology, Tbilisi State Medical University, Tbilisi, Georgia.

Corresponding author: Bakhtadze Sophio.

 E-mail: sophiabakhtadze@yahoo.com.

Abstract.

Impact of EEG biofeedback on event-related potentials (ERPs) in attention-deficit hyperactivity (ADHD) children.

Introduction: ADHD is one of the most widely spread condition of school aged children affecting 5% of children of this age. Diagnosis of ADHD increased and the rate had risen to 3.4% and it continues to rise. The core clinical signs of ADHD are inattention, restlessness and impulsivity. It is known that there are a numerous neurometric tools widely used for assessment of children with ADHD. According to various authors direct measures of attention are of two types:

1. Recording the alpha rhythm on the EEG and event-related potentials (ERP);
2. Tests of reaction time, continuous performance tests, paired associated learning, and tests of memorization;

The second one is evidence-based. As for the first one it is known that ERPs especially of those with later response reflect the process of mental effortfullness to select the appropriate behavior and accomplish decision making during the action of target stimulus. Thus selection of action as well as decision making is the most important points affected in ADHD children. Besides in recent years EEG biofeedback ( Neurofeedback) have become the evidence-based in the treatment of ADHD. Unfortunately the effectiveness of this approach on ERPs parameters is still unknown. Thus we aimed to study the changes of ERPs after neurofeedback therapy. Methods: We have examined 16 children with ADHD before- and after 3- sessions of neurofeedback therapy and 23 without treatment. Results: We have observed statistically significant improvement of parameters of later response like P300 in treated children compared with untreated ones whereas the treatment was non effective for earlier components of ERPs. Conclusions: Neurofeedback can affect on the process of selection of action and decision making by means of changing of P300 parameters in ADHDchildren.

**Introduction:** ADHD is one of the most widely spread condition of school aged children affecting 5% of children of this age [1]. Diagnosis of ADHD increased and the rate had risen to 3.4% and it continues to rise [2]. The core clinical signs of ADHD are inattention, restlessness and impulsivity. It is known that there are a numerous neurometric tools widely used for assessment of children with ADHD. According to various authors direct measures of attention are of two types [3]:

3. Recording the alpha rhythm on the EEG and event-related potentials (ERP);
4. Tests of reaction time, continuous performance tests, paired associated learning, and tests of memorization;

The last one is clinically proved and used in routine practice but as for the first one its validity is still ambiguous and has limited application in clinical practice especially of long latency event-related potentials (ERP). ERP studies is extremely important as it reflects multitude of cognitive factors such as attention, memory and language. Thus assessment of its components is very valuable in disorders accompanied with attention impairment including ADHD. The use of P3 for clinical assessment of various neurological diseases is recommended by International Federation of clinical neurophysiology [4]. ERP studies in ADHD started in 1970s with the work of Satterfield and his group [5].

ERP is recorded under certain conditions like using of certain modality stimulus. It is known that its latency increases as a categorization of the stimulus becomes more difficult [6], [7], [8] while others have reported that ERP latency remains unaffected by task difficulty [9]. The simplest way for auditory ERP to be recorded is during performing auditory oddball paradigm [10]. The core feature is to give series of stimuli with randomized consequence**.** These stimuli used for recording of ERP are divided into two ways: non- target and target which has to be recognized by patient. The ERP response recorded during recognition of relatively rare target stimulus differs from others by morphology as it appears with positive wave within latency range of 300 ms. This wave is called P3.

The response on target stimulus can be divided into two ways:

1. Earlier response P1-N1-P2 connected to simple auditory stimulus.
2. Relatively later N2-P3-N3-complex related to cognition like recognition and differentiation of stimulus and decision making which patient can express by pushing the button when hearing and recognizing target stimulus.

The earlier component like N1 - a first discernible peak is a negativity around 80 ms with a frontocentral maximum present for both rare and frequent tones [4]. It generally thought to represent the initial extraction of information from sensory analysis of the stimulus [11] and represents the correct recognition of stimulus [12].

The most significant component of later response is P3. It is proposed that amplitude and latency of P3 reflect the extent and timing of a distribution-specific cognitive process, while scalp distribution is indicative of which cognitive process is activated during the performance of a task [13].

The P3 reflects multiple cognitive processes, specifically attentional resource allocation [14]. P3 amplitude is thought to be a reflection of effortfulness of the stimulus response and the intensity of processing whereas P3 latency is taken as a reflection of the speed of information processing [15]. The assessment of P3 is important in ADHD children for several reasons: The first it is known that most important aspect of executive functioning frequently affected in ADHD children is engagement operation [10]. This process from neurophysiological point of view is associated with the activation of cortical and subcortical structures that are involved in execution of the selected action which is disturbed in ADHD children. From psychological, functional point of view the engagement operation is associated with combining all brain resources for the action to be accomplished [10]. This operation is manifested in the P3 components. And the second according to Desmedt and Debecker's hypothesis the occurrence of P3 corresponds to the termination of the decision-making process which is taking place during the categorization of the target stimulus [16].

There are several studies concerning clinical values of ERP in children with ADHD. In spite of enormous studies in this field the unified consensus about morphology and clinical value of P3 in ADHD children is still absent. According to various authors P3 in ADHD does not differ from that of control group [17] whereas others detected significant changes in amplitude and latency of P3 in ADHD vs controls [10], [18]. According to Alexander and colleagues [19] P3 can serve as a valid marker of attention disorders in ADHD and predictor of pharmacological treatment [20], [21], [22].

As for non pharmacological intervention less is known about impact of EEG Biofeedback (neurofeedback-NF) on P3 characteristics although NF becomes evidence based treatment option for ADHD in recent years. According to evidence-based practice in biofeedback and neurofeedback recommended by international society for mind-body research, health care and education, the efficacy of NF in ADHD has satisfied the upper level of efficacy ( level 4 efficacy) [23].. Such popularity of NF can be explained by the fact that stimulants frequently used for the treatment of ADHD can cause various side effects including growth suppression. In addition, estimates indicate that as many as 30% of children with ADHD either do not respond to stimulant treatment or cannot tolerate the treatment secondary to side effects. This has lead to the consideration of treatment with both nonstimulant medications as well as alternative therapies, including diet, iron supplementation and NF [24]. Only several papers are devoted to impact of NF on ERP measures [10], [25]. According to Wangler and colleagues [25], the decrease of P3 latency was observed after NF therapy but according to these authors these evidence is less valid as 18 units for a single NF protocol might have been to small to obtain specific ERP effects. Besides this study is not consistent because of absence of comparative analysis as it did not include non - treated ADHD children as a control group.

Thus assessment of ERP parameters before- and after NF treatment is extremely important in ADHD children.

**Materials and methods:** We prospectively studied 39 children with ADHD of combined subtype (ADHDcom) without any kind of pharmacological treatment. Age range 10-12 years. All children met the diagnostic criteria for ADHDcom according to the DSM-IV (American Academy of Pediatrics: Clinical Practice Guideline: Diagnosis and Evaluation of the Child With Attention-Deficit/Hyperactivity Disorder. Pediatrics 2000; 105 (5): 1158-1170). ADHD was diagnosed by board certified experienced neurologist and a paediatric neurologist based on the child's detailed neurological and neuropsychological examination and a detailed caregiver interview using Conners parent rating scale [26]. This group was divided into two subgroups: The first ADHDcom-1 (16 children) were children where NF treatment was carried out and the second subgroup of ADHDcom-2 (23 children) were non treated children. Such distribution of patients can be explained by the fact that P3 [27] as well as N1 **[28]** can exhibit habituation by repetition after certain time interval.

We obtained informed written consent from parents or guardians of all children. The study protocol was approved by the Biomedical Research Ethic Committee in Tbilisi State Medical University.

A neuropsychological assessment was conducted in all group of children, including IQ testing by the Wechsler Intelligence Scale for Children-Revised (WISC-R). The inclusion criteria were IQ>70**.** All other children assessed were determined to be intellectually disabled and were excluded from study.

ERP was also carried out in both group of children .The parameters of ERP assessed were polarity (positive or negative), timing (latency) and amplitude. Omission (misses) and commission (false alarms) errors were also investigated as they are thought to index impulsivity [29]. EEG was recorded using with silver plated electrodes placed at area according to the international 10-20 system. We have chosen only Cz-A1 derivation for the analysis for two reasons:

1. It is known that amplitude of N1 as well as P3 at target stimuli is the largest in the midline [4], [18]. In order to avoid interregional effect frequently observed in ERP studies as it was out of scope of our study.

Cap electrodes were referenced to linked ears. Impedance was kept below 5 kµ for the cap electrodes. The subject was grounded by the cap ground electrodes located midway between Fpz and Fz. EEG was amplified 20 000 times with a bandpass down 3dB at 0.1 and 25 Hz and was sampled through a 12-bit analog-to-digital converter at 200 Hz. In the laboratory each child was familiarized with the testing equipment and procedure. All tasks were completed in a sound-attenuated air-conditioned room. Seated on a comfortable chair children were presented with auditory stimuli binaurally through headphones. Conditions of stimulations were:

- Stimulation-binaural
- Duration of stimulus-50ms.

- Intensiveness - 80Db
- Interstimulus period-1 second
- Tone frequency:

- For target stimulus-2000Hz, reliability 20-30%,
- For non-target stimulus-1000Hz, reliability 70-80%.

Non-target and target stimuli were given in an absolutely randomized way 1-2 target stimulus appears after each 5 non-target stimuli.

For more refinement of study we choose only two components of ERP: N1 as an earlier electrical correlate of cortical activity reflecting correct choice of task and P3 as a later one reflecting complex act of executive functions including attention.

NF was carried out in ADHDcom-1 children according to protocol previously developed by Lubar [30] It was conducted over a period of 6 weeks (30 sessions, five training per week). Training was divided into two phases (15 sessions in each phase): in the first phase children were trained to enhance the amplitude of SMR (12-15) Hz and decrease the amplitude of theta activity (4-7Hz); In the second phase beta/theta training was used which required children to decrease the amplitude of theta waves (4-7Hz) and reward their beta 1 waves (15-18 Hz). During the first phase EEG was recorded from Cz, with reference placed on the left earlobe and ground electrode on the right earlobe. During the second phase EEG was recorded from Fz-Pz derivations with the same localization of the reference and ground electrodes. It is known that SMR enhancement reduces hyperactivity problems. As for beta/theta training the basic assumption guiding this approach is that suppressing theta activity diminishes attention problems [31]. Each session was subdivided into 2 min periods (that were gradually increased up to 10 min). During the training sessions children were attempting to work with puzzles on the screen, solve mathematical problems and maintain a state relaxation by watching cartoons.

Data were analyzed by different descriptive and inferential statistical methods using SPSS 10.0. The 2 groups × 2 conditions repeated measures ANOVA was used to determine the effect of treatment on ERPs for the study groups. The correlation analysis was made by the Spearman's rho and the Pearson correlations. The independent-sample t test was used to make the between group comparisons. The nonparametric Mann Whitney U test was used to make between group comparisons for independent samples. The nonparametric Mann Whitney U test was used to compare ADHD groups with and without treatment.

## Results:

Case-control study of assessing ERPs was performed by forming groups of ADHD children without treatment and ADHD children with treatment. Table 1 presents the percentage distribution of study groups, mean age and means of latency $N_1$, amplitude $N_1$, latency $P_3$, amplitude $P_3$, of omission and commission error rates before and after treatment (Figure 1, Figure 2).

Table 1. The percentage distribution of the study groups, mean age and means of latency, amplitude, and of error rates before and after treatment.

| Group N (%) | Age M (σ; min-max) | Latency $N_1$ | | Amplitude $N_1$ | | Latency $P_3$ | | Amplitude $P_3$ | | Omission errors | | Commission errors | |
|---|---|---|---|---|---|---|---|---|---|---|---|---|---|
| | | Before treatment | After treatment | Before treatment | After treatment | Before treatment | After treatment | Before treatment | After treatment | Before treatment | After treatment | Before treatment | After treatment |
| ADHD with treatment 16 (41 %) | 10.63 (.81; 10-12) | 115.19 (15.93; 83-133) | 113.81 (16.19; 81-132) | 5.31 (2.7; 1-10) | 5.38 (2.68; 2-10) | 465.94 (33.73; 401-523) | 417.31 (22.51; 357-441) | 6.87 (2.03; 3-10) | 9.81 (1.87; 7-13) | 5.13 (1.59; 2-8) | 2.38 (.89; 1-4) | 4.5 (2.31; 1-8) | .88 (.89; 0-3) |
| ADHD without treatment 23 (59 %) | 10.57 (.73; 10-12) | 116.78 (15.22; 85-135) | 115.61 (16.68; 83-141) | 3.61 (1.75; 1-7) | 4.13 (1.82; 1-8) | 445.26 (51.26; 354-501) | 420.13 (45.4; 343-487) | 6.39 (1.67; 3-9) | 7.3 (1.69; 3-10) | 5.39 (2.13; 1-9) | 4.7 (1.79; 2-7) | 4.3 (1.61; 2-8) | 4.87 (2.91; 0-9) |

Figure 1. The percentage distribution and the mean age of the study groups.

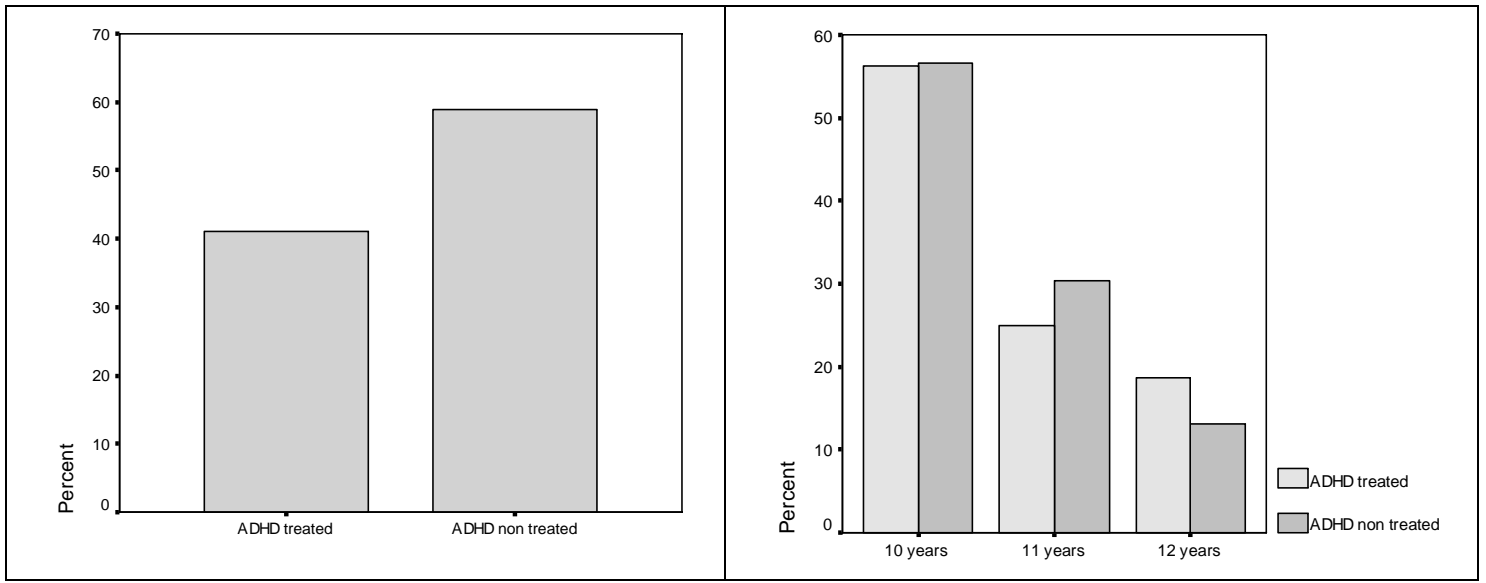

Figure 2. The mean performance of latency, amplitude, and of error rates before and after treatment for study groups.

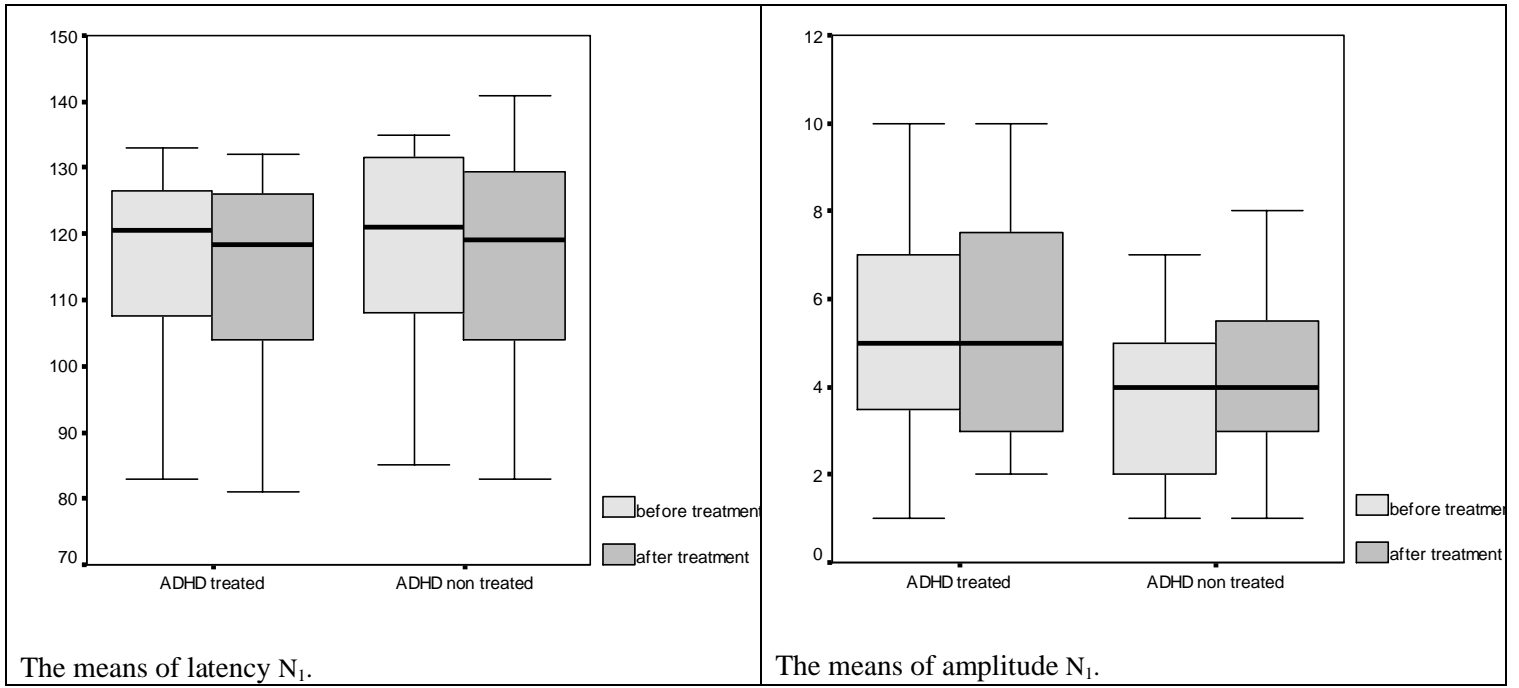

The means of latency $N_1$.

The means of amplitude $N_1$.

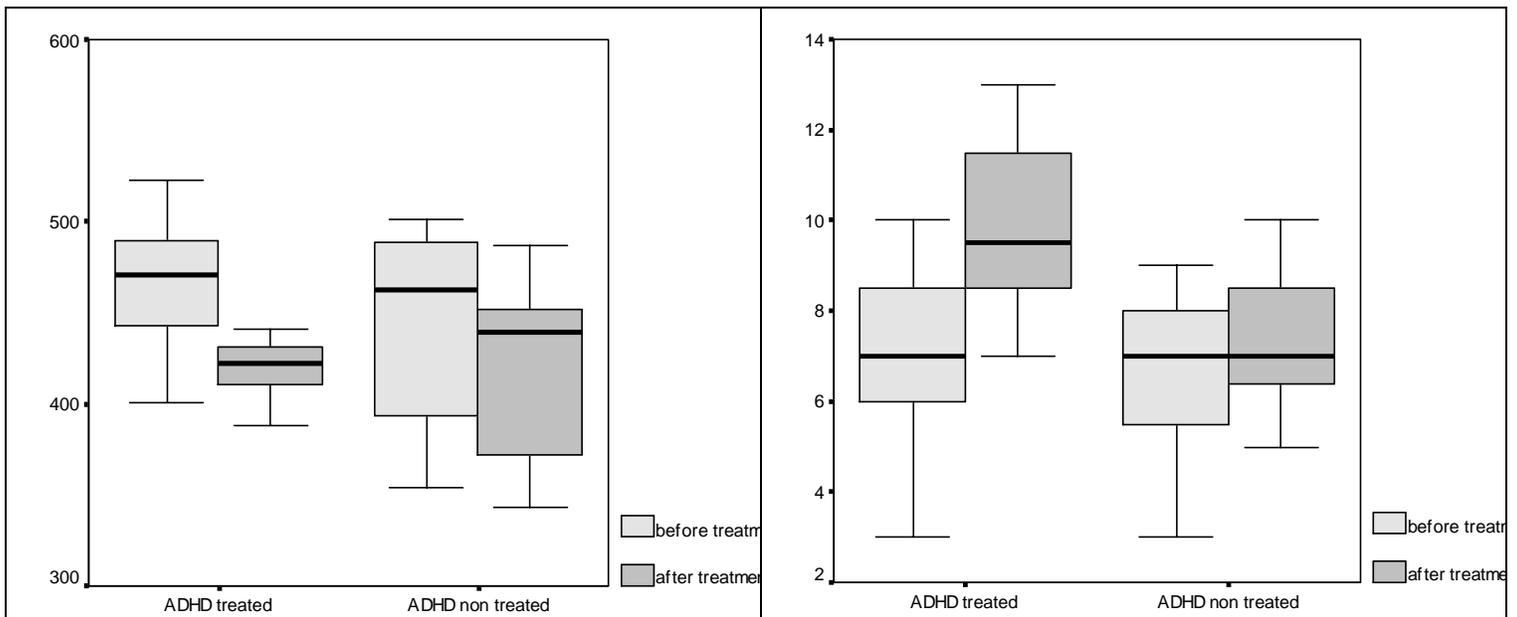

| The means of latency P$_3$. | The means of amplitude P$_3$. |

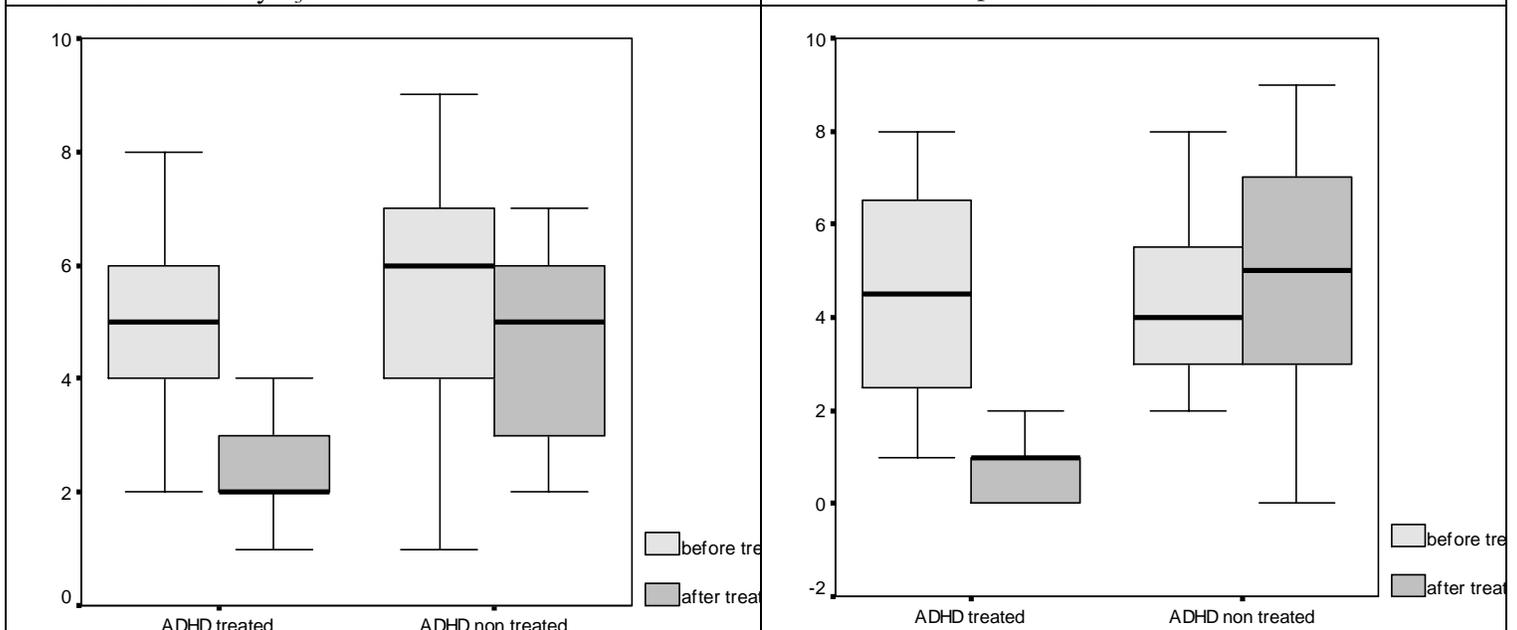

| The means of omission error rates. | The means of commission error rates. |

**Results:**

The mean amplitude N$_1$ before treatment was significantly higher t(23.676)=2.219, p<.036 for the ADHD treated group (mean=5.31, SD=2.7) compared to the ADHD non treated group (mean=3.61, SD=1.75) (Figure 2). The mean amplitude P$_3$ after treatment was significantly higher t(47)=4.376, p<.0001 for the ADHD treated group (mean=9.81, SD=1.87) compared to the ADHD non treated group (mean=7.3, SD=1.69). The mean omission errors rate after treatment was significantly higher t(33.979)=-5.338, p<.0001 for the ADHD non treated group (mean=4.7, SD=1.79) compared to the ADHD treated group (mean=2.38, SD=.89). The mean commission errors rate after treatment was significantly higher t(27.518)=-6.18, p<.0001 for the ADHD non treated group

(mean=4.87, SD=2.91) compared to the ADHD treated group (mean=.88, SD=.89). The same results were obtained for these variables using the nonparametric Mann Whitney U test.

The Wilcoxon matched-pairs, signed rank test for the group of ADHD children without treatment showed that the difference between the median latency $N_1$ before treatment (M=116.78, SD=15.22) and median latency $N_1$ after time interval (M=115.61, SD=16.68) was significant beyond .05 level: Z=-2.2, p<.028. The median latency $N_1$ at the first time was significantly higher than median latency $N_1$ after time interval. The difference between the median amplitude $N_1$ before treatment (M=3.61, SD=1.75) and median amplitude $N_1$ after time interval (M=4.13, SD=1.82) was significant beyond .05 level: Z=-2.546, p<.011. The median latency $N_1$ after time interval was significantly higher than median amplitude $N_1$ before treatment. The difference between the median latency $P_3$ at the first time (M=445.26, SD=51.26) and median latency $P_3$ after time interval (M=420.13, SD=45.4) was significant beyond .001 level: Z=-4.168, p<.0001. The median latency $P_3$ at the first time was significantly higher than median latency $P_3$ after time inetraval. The difference between the median amplitude $P_3$ at the first time (M=6.39, SD=1.67) and median amplitude $P_3$ after time interval (M=7.3, SD=1.69) was significant beyond .001 level: Z=-3.493, p<.0001. The median amplitude $P_3$ after time interval was significantly higher than median amplitude $P_3$ at the first time. The difference between the median omission error rate at the first time (M=5.39, SD=2.13) and median omission error rate after time interval (M=4.7, SD=1.79) was significant beyond .05 level: Z=-2.054, p<.04. The median omission error rate at the first time was significantly higher than median omission error rate after time interval.

The Wilcoxon matched-pairs, signed rank test for the group of ADHD children with treatment showed that the difference between the median latency $N_1$ before treatment (M=115.19, SD=15.93) and median latency $N_1$ after treatment (M=113.81, SD=16.19) was significant at one tail Z=-1.898, p<.058. The median latency $N_1$ before treatment was significantly higher than median latency $N_1$ after treatment. The difference between the median latency $P_3$ before treatment (M=465.94, SD=33.73) and median latency $P_3$ after treatment (M=417.31, SD=22.51) was significant beyond .001 level: Z=-3.517, p<.0001. The median latency $P_3$ before treatment was significantly higher than median latency $P_3$ after treatment. The difference between the median amplitude $P_3$ before treatment (M=6.87, SD=2.03) and median amplitude $P_3$ after treatment (M=9.81, SD=1.87) was significant beyond .001 level: Z=-3.546, p<.0001. The median amplitude $P_3$ after treatment was significantly higher than median amplitude $P_3$ before treatment. The difference between the median omission error rate before treatment (M=5.13, SD=1.59) and median omission error rate after treatment (M=2.38, SD=.89) was significant beyond .001 level: Z=-3.539, p<.0001. The median omission error rate before treatment was significantly higher than median omission error rate after treatment. The difference between the median commission error rate before treatment (M=4.5, SD=2.31) and median commission error rate after treatment (M=.88, SD=.89) was significant beyond .001 level: Z=-3.426, p<.001. The median commission error rate before treatment was significantly higher than median commission error rate after treatment.

To determine the effect of treatment on latency $N_1$, the data from the latency $N_1$ before treatment and latency $N_1$ after treatment were entered separately into a 2 (group) × 2 (condition – latency $N_1$ before treatment vs. latency $N_1$ after treatment) ANOVA. The ANOVA showed a significant effect of condition $F(1,37)=6.282$, $MSE=4.879$, $p<.017$, partial eta squared $=.145$ represents a large effect size. The effects of group ($F<1$) and interaction between group and condition ($F<1$) were not significant. The effect of age was significant $F(1,36)=182.252$, $MSE=86$, $p<.0001$, partial eta squared $=.835$ represents a large effect size, but the interaction between age and condition wasn't significant ($F<1$). The taking of age as covariate change relationships between the study variables and the effect of condition became not significant.

To determine the effect of treatment on amplitude $N_1$, the data from the amplitude $N_1$ before treatment and amplitude $N_1$ after treatment were entered separately into a 2 (group) × 2 (condition – amplitude $N_1$ before treatment vs. amplitude $N_1$ after treatment) ANOVA. The ANOVA showed a significant effect of condition $F(1,37)=4.829$, $MSE=.333$, $p<.034$, partial eta squared $=.145$ represents a large effect size, and of group $F(1,37)=4.399$, $MSE=9.323$, $p<.043$, partial eta squared $=.106$ represents a close to medium effect size. The effects of interaction between group and condition was significant at one tail $F(1,37)=2.984$, $MSE=.333$, $p<.092$. The effect of age was significant $F(1,36)=23.655$, $MSE=5.782$, $p<.0001$, partial eta squared $=.397$ represents a large effect size, but the interaction between age and condition wasn't significant ($F<1$) (Figure 3). The taking of age as covariate didn't change relationships between the study variables.

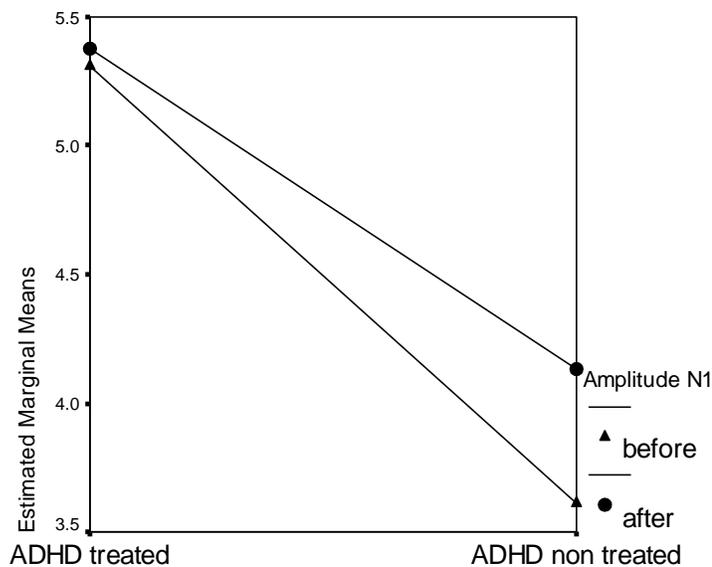

Figure 3. The means of amplitude $N_1$ before and after treatment for study groups.

To determine the effect of treatment on latency $P_3$, the data from the latency $P_3$ before treatment and latency $P_3$ after treatment were entered separately into a 2 (group) × 2 (condition – latency $P_3$ before treatment vs. latency $P_3$ after treatment) ANOVA. The ANOVA showed a significant effects of condition $F(1,37)=223.69$, $MSE=114.735$, $p<.0001$, partial eta squared $=.858$ represents a large effect size and of interaction between group and condition $F(1,37)=22.698$, $MSE=114.735$, $p<.0001$, partial eta squared $=.38$ represents a large effect size. The effect of group was not significant (F<1). The ANOVA showed a significant effects effect of age F(1,36)=40.62, *MSE*=1612.812, p<.0001, partial eta squared =.53 represents a large effect size, and interaction between age and condition F(1,36)=26.96, *MSE*=67.427, p<.0001, partial eta squared =.428 represents a large effect size (Figure 4). The taking of age as covariate didn't change relationships between the study variables.

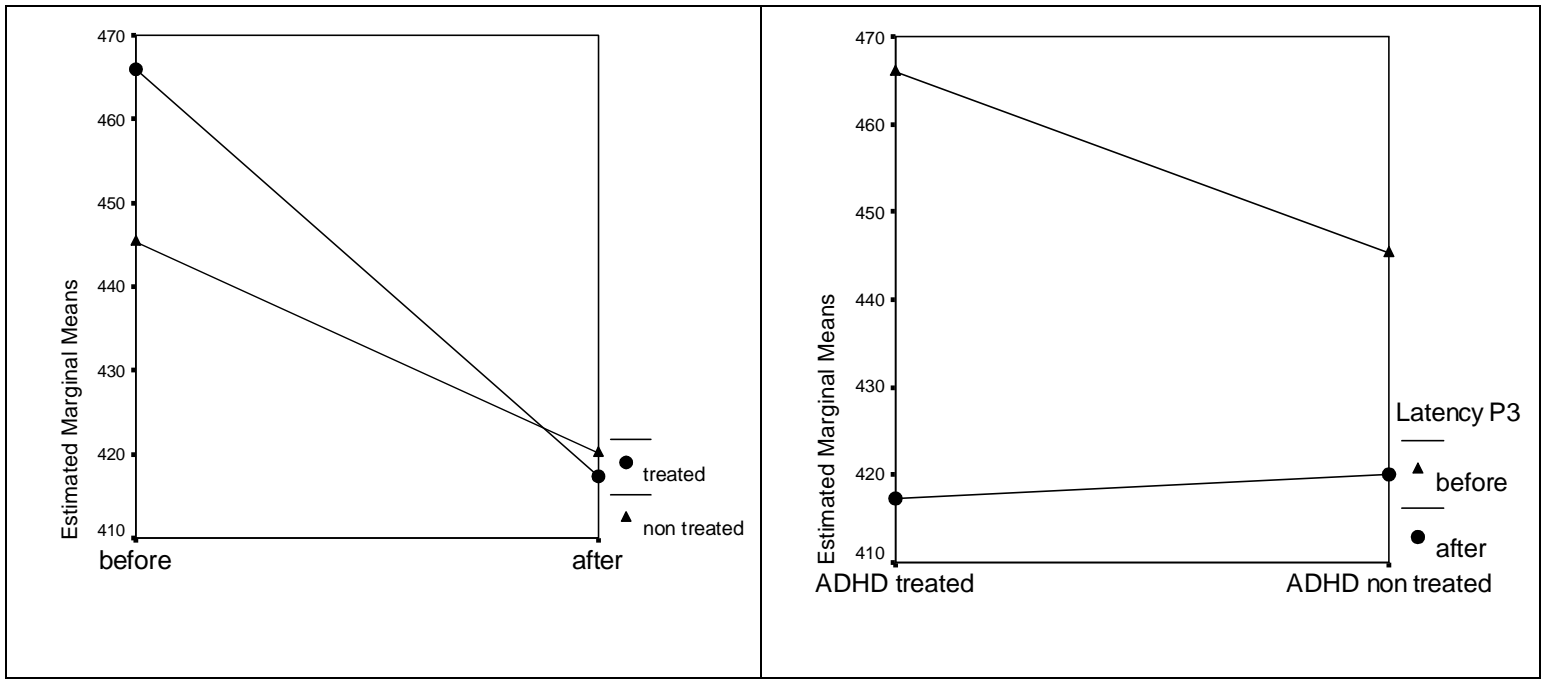

Figure 4. The means of latency $P_3$ before and after treatment for study groups.

To determine the effect of treatment on amplitude $P_3$, the data from the amplitude $P_3$ before treatment and amplitude $P_3$ after treatment were entered separately into a 2 (group) × 2 (condition – amplitude $P_3$ before treatment vs. amplitude $P_3$ after treatment) ANOVA. The ANOVA showed a significant effects of condition F(1,37)=157.263, *MSE*=.443, p<.0001, partial eta squared =.81 represents a large effect size; of group F(1,37)=7.07, *MSE*=6.008, p<.012, partial eta squared =.16 represents a large effect size, and of interaction between group and condition F(1,37)=44.044, *MSE*=.443, p<.0001, partial eta squared =.543 represents a large effect size. The effect of age was significant F(1,36)=7.721, *MSE*=5.085, p<.009, partial eta squared =.177 represents a large effect size but the interaction between age and condition wasn't significant (F<1) (Figure 5). After taking of age as covariate the effect of condition became not significant but didn't change other relationships between the study variables.

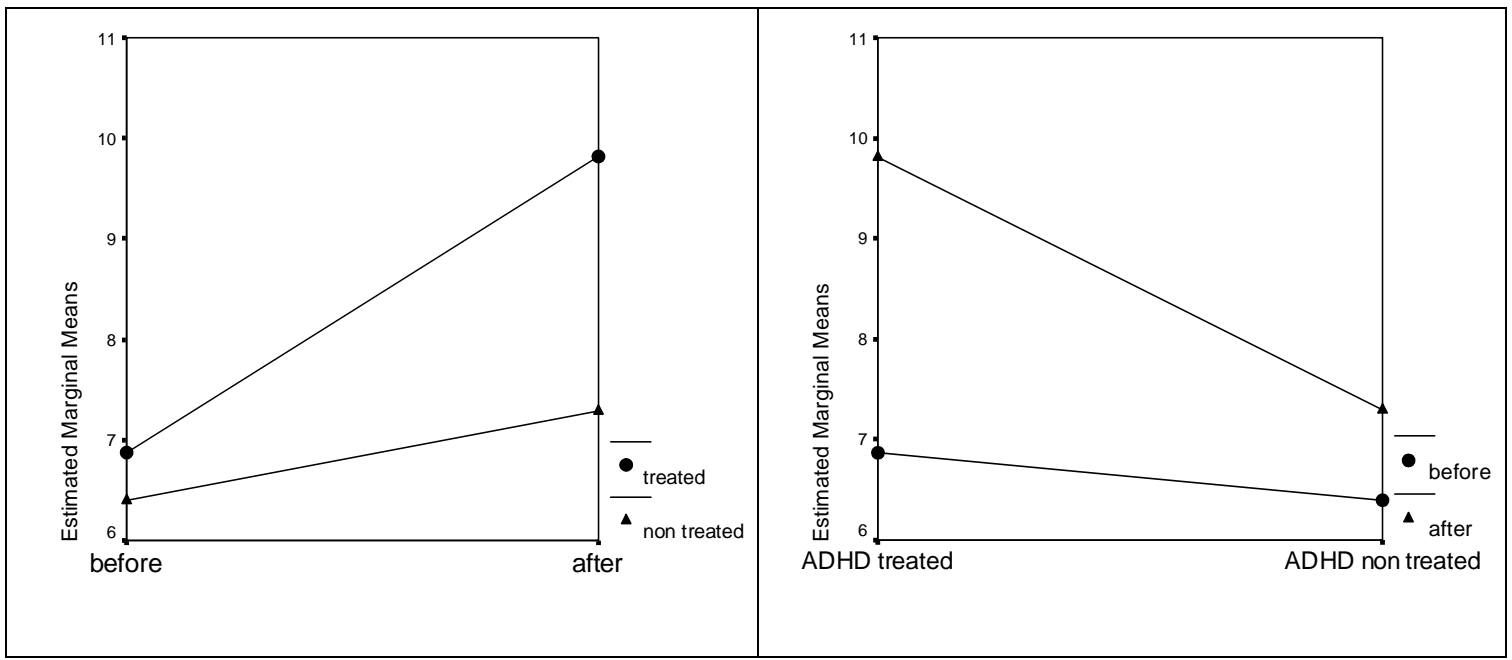

Figure 5. The means of amplitude $P_3$ before and after treatment for study groups.

To determine the effect of treatment on omission error rate, the data from the omission error rate before treatment and omission error rate after treatment were entered separately into a 2 (group) × 2 (condition – omission error rate before treatment vs. omission error rate after treatment) ANOVA. The ANOVA showed a significant effects of condition $F(1,37)=54.634$, *MSE*=1.025, p<.0001, partial eta squared =.596 represents a large effect size; of group $F(1,37)=6.422$, *MSE*=4.917, p<.016, partial eta squared =.148 represents a large effect size and of interaction between group and condition $F(1,37)=19.421$, *MSE*=1.025, p<.0001, partial eta squared =.344 represents a large effect size. The ANOVA showed a significant effects effect of age $F(1,36)=61.727$, *MSE*=1.861, p<.0001, partial eta squared =.632 represents a large effect size, and interaction between age and condition $F(1,36)=9.672$, *MSE*=.831, p<.004, partial eta squared =.212 represents a large effect size (Figure 6). The taking of age as covariate didn't change relationships between the study variables.

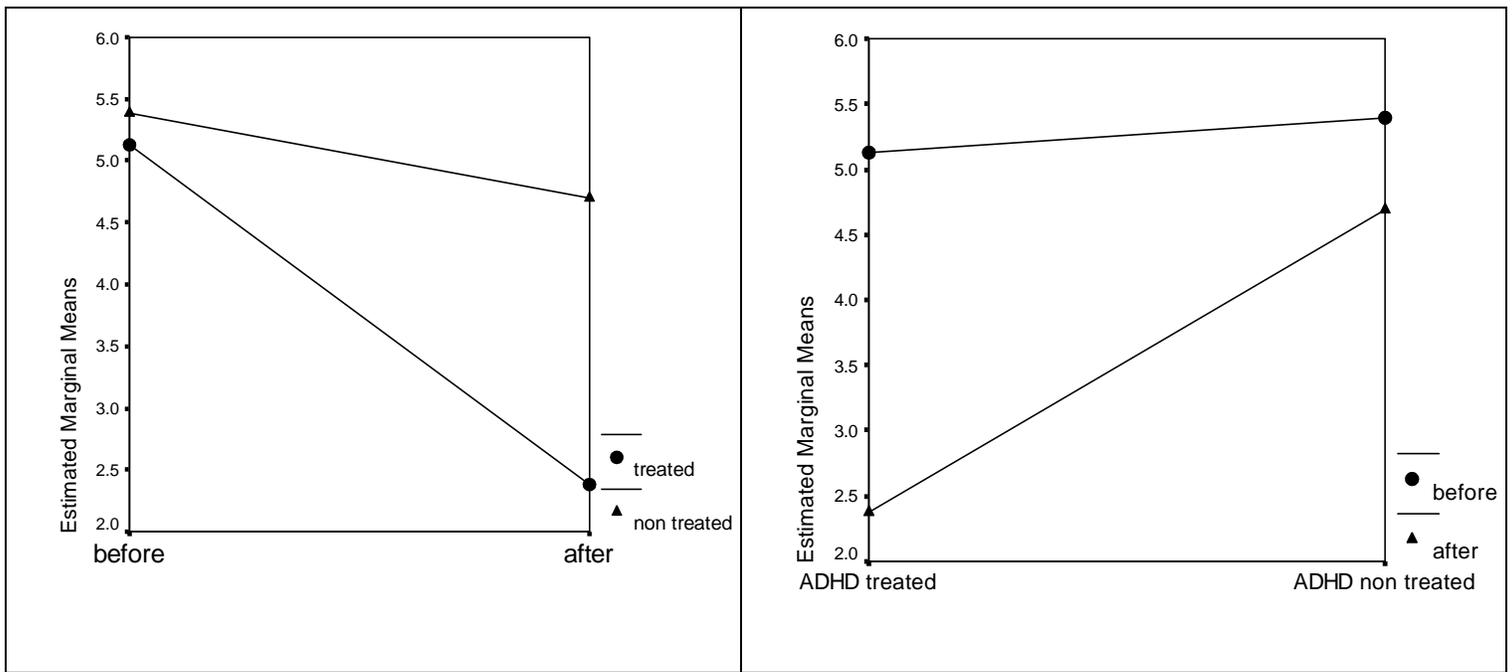

Figure 6. The means of omission error rate before and after treatment for study groups.

To determine the effect of treatment on commission error rate, the data from the commission error rate before treatment and commission error rate after treatment were entered separately into a 2 (group) × 2 (condition – commission error rate before treatment vs. commission error rate after treatment) ANOVA. The ANOVA showed a significant effects of condition F(1,37)=20.004, *MSE*=2.208, p<.0001, partial eta squared =.351 represents a large effect size; of group F(1,37)=9.937, *MSE*=6.852, p<.003, partial eta squared =.212 represents a large effect size, and of interaction between group and condition F(1,37)=37.515, *MSE*=2.208, p<.0001, partial eta squared =.503 represents a large effect size. The effects of age (F<1) and interaction between age and condition were not significant (F<1) (Figure 7). After taking of age as covariate the effect of condition became not significant but didn't change other relationships between the study variables.

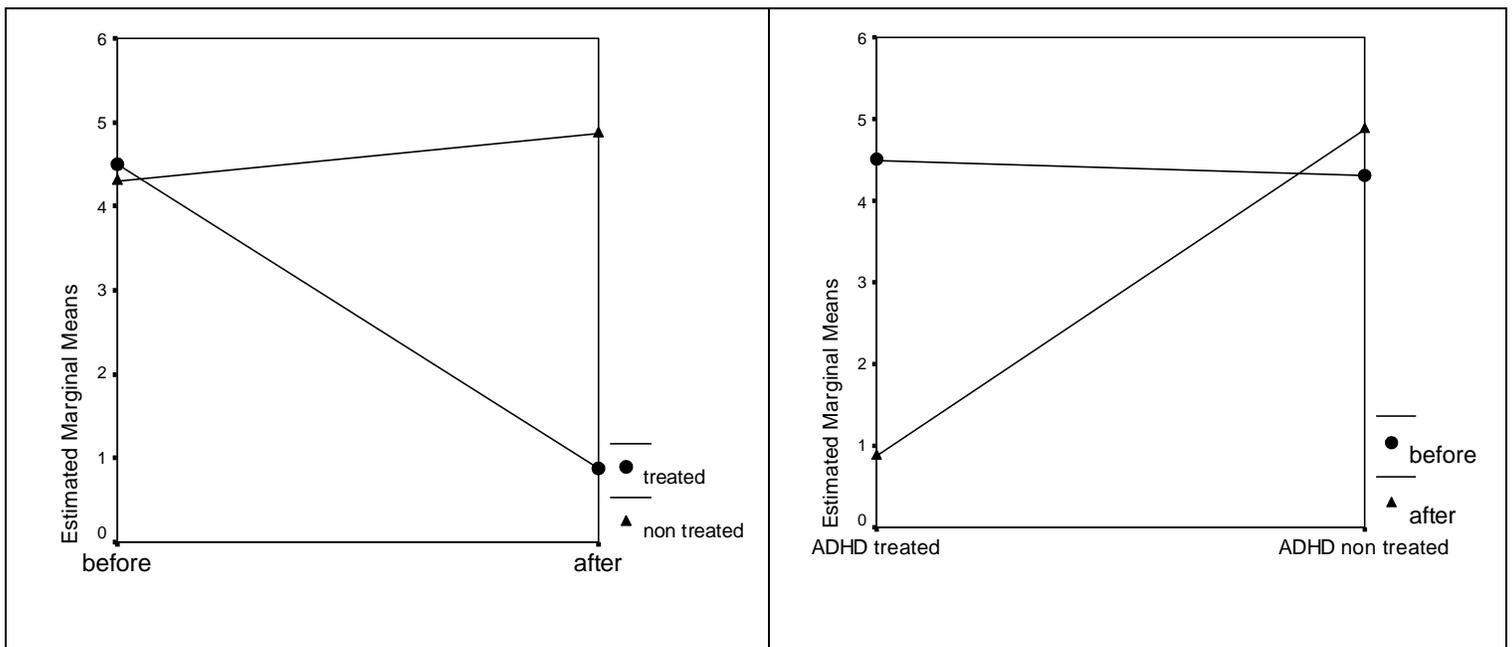

Figure 7. The means of commission error rate before and after treatment for study groups.

To determine the effect of treatment on latency, the data from the latency $N_1$ before treatment and latency $P_3$ before treatment were entered separately into a 2 (group) × 2 (condition – latency $N_1$ before treatment vs. latency $P_3$ before treatment) ANOVA. The ANOVA showed a significant effect of condition $F(1,37)=3712.44$, $MSE=586.307$, $p<.0001$, partial eta squared $=.99$ represents a large effect size. The effect of group was not significant. The effect of interaction between group and condition was significant at one tail $F(1,37)=3.992$, $MSE=586.307$, $p<.053$. The effects of age $F(1,36)=90.941$, $MSE=489.096$, $p<.0001$, partial eta squared $=.716$ represents a large effect size, and interaction between age and condition $F(1,36)=19.768$, $MSE=338.991$, $p<.0001$, partial eta squared $=.354$ represents a large effect size, were significant (Figure 8). The taking of age as covariate change relationships between the study variables and revealed the significant effect of interaction between group and condition $F(1,36)=6.902$, $MSE=388.991$, $p<.013$, partial eta squared $=.161$ represents a large effect size.

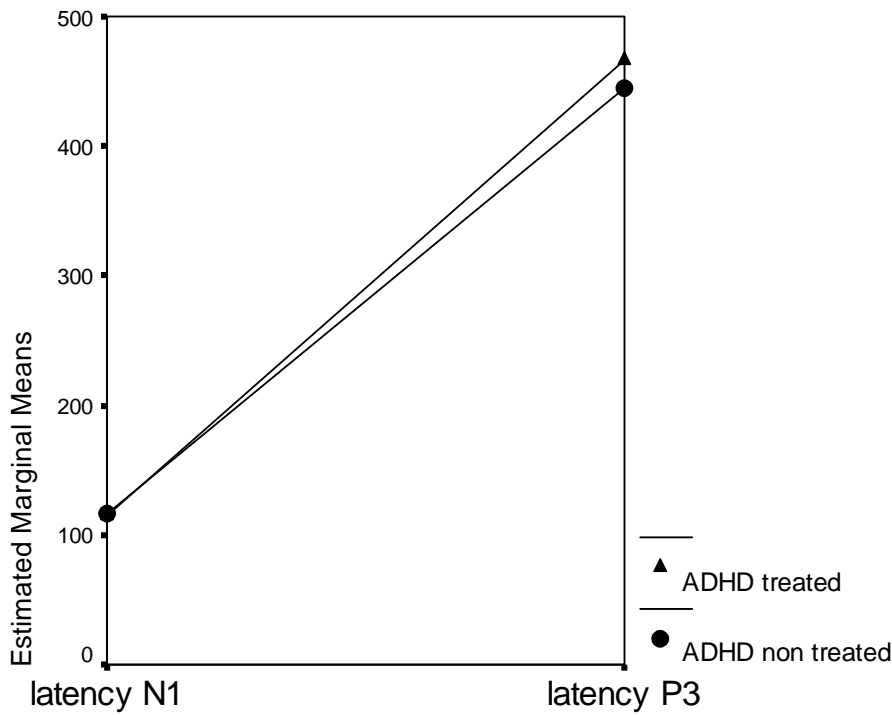

Figure 8. The means of latency $N_1$ and $P_3$ before treatment for study groups.

To determine the effect of treatment on amplitude, the data from the amplitude $N_1$ before treatment and amplitude $P_3$ before treatment were entered separately into a 2 (group) × 2 (condition – amplitude $N_1$ before treatment vs. amplitude $P_3$ before treatment) ANOVA. The ANOVA showed a significant effects of condition $F(1,37)=53.222$, $MSE=1.674$, $p<.0001$, partial eta squared $=.59$ represents a large effect size, and interaction between group and condition $F(1,37)=4.196$, $MSE=1.674$, $p<.048$, partial eta squared $=.102$ represents a medium effect size. The effect of group was significant at one tail $F(1,37)=6.438$, $MSE=3.507$, $p<.069$. The effect of age was significant $F(1,36)=18.78$, $MSE=4.349$, $p<.0001$, partial eta squared $=.343$ represents a large effect size but the interaction between age and condition was significant at one tail $F(1,36)=3.242$, $MSE=1.578$, $p<.08$ (Figure 9). The taking of age as covariate didn't change relationships between the study variables.

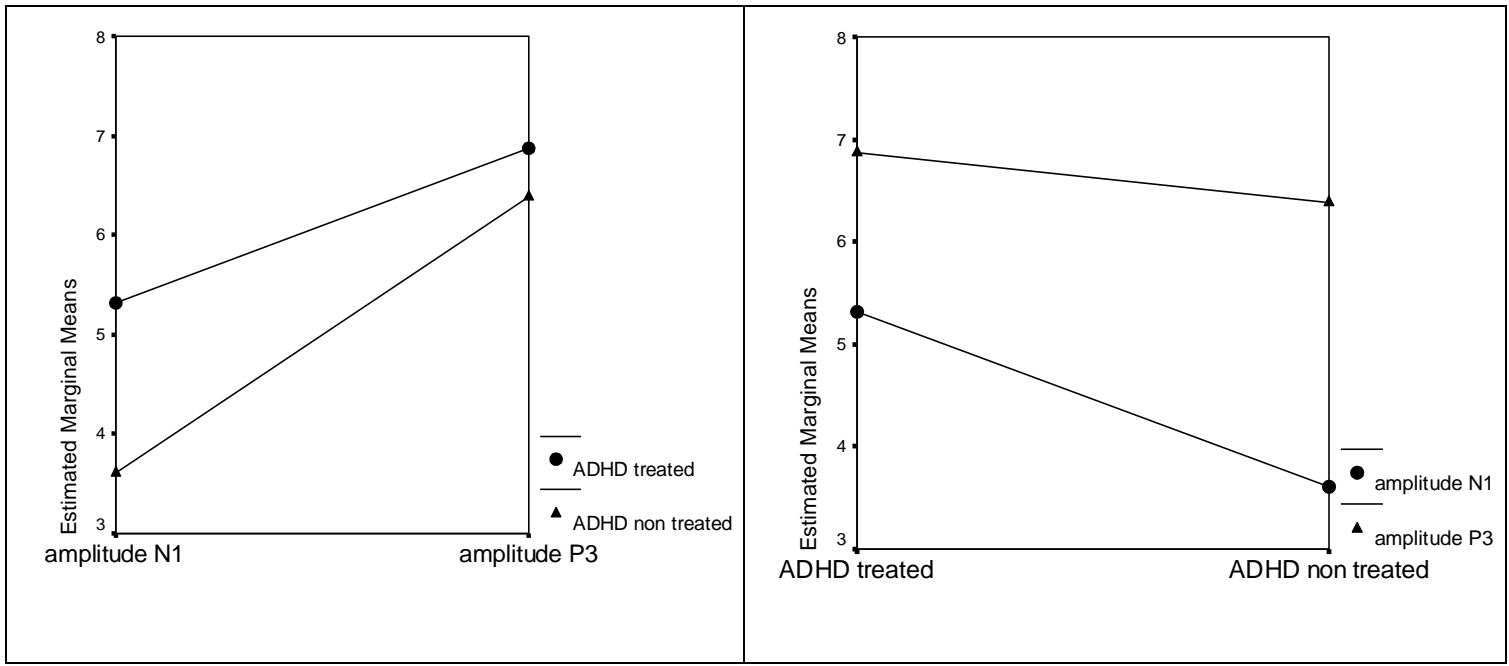

Figure 9. The means of amplitude $N_1$ and $P_3$ before treatment for study groups.

To determine the effect of treatment on latency, the data from the latency $N_1$ after treatment and latency $P_3$ after treatment were entered separately into a 2 (group) × 2 (condition – latency $N_1$ after treatment vs. latency $P_3$ after treatment) ANOVA. The ANOVA showed a significant effect of condition $F(1,37)=3935.793$, $MSE=443.159$, $p<.0001$, partial eta squared $=.991$ represents a large effect size. The effects of group and interaction between group and condition were not significant. The effect of age $F(1,36)=59.26$, $MSE=489.116$, $p<.0001$, partial eta squared $=.622$ represents a large effect size, and interaction between age and condition $F(1,36)=4.592$, $MSE=403.941$, $p<.039$, partial eta squared $=.113$ represents a large effect size, were significant. The taking of age as covariate didn't change relationships between the study variables.

To determine the effect of treatment on amplitude, the data from the amplitude $N_1$ after treatment and amplitude $P_3$ after treatment were entered separately into a 2 (group) × 2 (condition – amplitude $N_1$ after treatment vs. amplitude $P_3$ after treatment) ANOVA. The ANOVA showed a significant effects of condition $F(1,37)=138.457$, $MSE=1.97$, $p<.0001$, partial eta squared $=.789$ represents a large effect size, and group $F(1,37)=11.078$, $MSE=6.026$, $p<.002$, partial eta squared $=.23$ represents a large effect size. The effect of interaction between group and condition was significant at one tail $F(1,37)=3.877$, $MSE=1.97$, $p<.056$. The effect of age $F(1,36)=20.005$, $MSE=3.981$, $p<.0001$, partial eta squared $=.357$ represents a large effect size, and interaction between age and condition $F(1,36)=5.747$, $MSE=1.746$, $p<.022$, partial eta squared $=.138$ represents a large effect size, were significant (Figure 10). The taking of age as covariate change relationships between the study variables and revealed the significant effect of interaction between group and condition F(1,36)=4.775, *MSE*=1.746, p<.035, partial eta squared =.117 represents a large effect size.

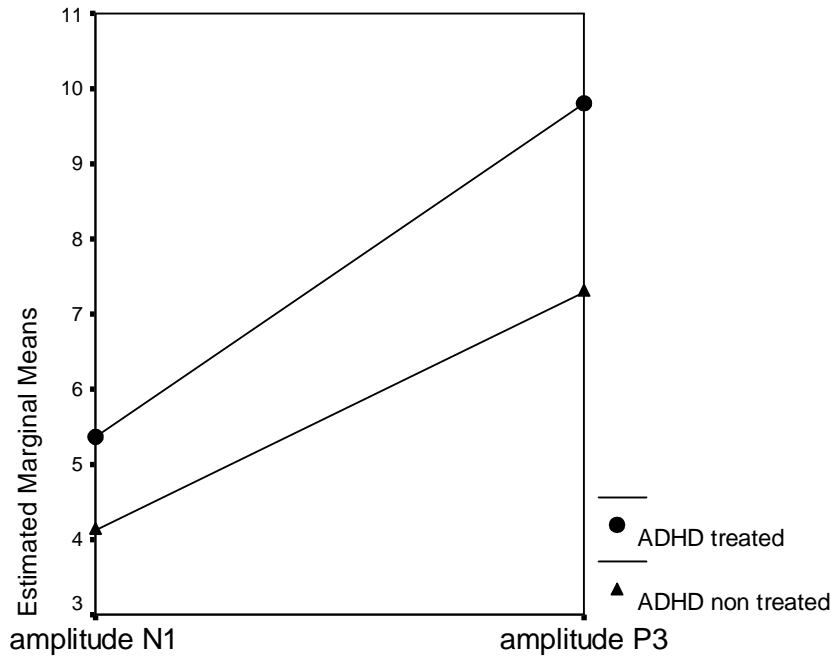

Figure 10. The means of amplitude $N_1$ and $P_3$ after treatment for study groups.

The correlation analysis didn't show significant relationships between commission error rates and the study variables. The significant correlations were not found between omission and commission error rates. The correlation between the omission error rate before treatment and commission error rate before treatment was significant at one tail $r_s$=.45, p<.083 for the group of ADHD children with treatment but not for the group of ADHD children without treatment. The significant correlation was found between the commission error rates before and after treatment r=.36, p<.025 (for the group of ADHD children with treatment correlation was significant at one tail $r_s$=.46, p<.075; for the group of ADHD children without treatment significant correlation was found $r_s$=.71, p<.0001).

The high significant correlations were found between age and latency $N_1$ before treatment r=-.92, p<.0001, amplitude $N_1$ before treatment r=.57, p<.0001, latency $N_1$ after treatment r=-.90, p<.0001, amplitude $N_1$ after treatment r=.62, p<.0001, latency $P_3$ before treatment r=-.74, p<.0001, amplitude $P_3$ before treatment r=.43, p<.006, latency $P_3$ after treatment r=-.66, p<.0001, amplitude $P_3$ after treatment r=.33, p<.04, omission error rate before treatment r=-.82, p<.0001 and omission error rate after treatment r=.51, p<.001 (for the group of ADHD children with treatment significant correlations were found between age and latency $N_1$ before treatment $r_s$=-.76, p<.001, amplitude $N_1$ before treatment $r_s$=.85, p<.0001, latency $N_1$ after treatment $r_s$=-.75, p<.001, amplitude $N_1$ after treatment $r_s$=.86, p<.0001, latency $P_3$ before treatment $r_s$=-.84, p<.0001, amplitude $P_3$ before treatment $r_s$=.69, p<.003, latency $P_3$ after treatment $r_s$=-.65, p<.006, amplitude $P_3$ after treatment $r_s$=.63, p<.009, omission error rate before treatment $r_s=-.72$, $p<.002$ and omission error rate after treatment $r_s=-.69$, $p<.003$; for the group of ADHD children without treatment significant correlations were found between age and latency $N_1$ before treatment $r_s=-.89$, $p<.0001$, amplitude $N_1$ before treatment $r_s=.41$, $p<.05$, latency $N_1$ after treatment $r_s=-.89$, $p<.0001$, amplitude $N_1$ after treatment $r_s=.48$, $p<.022$, latency $P_3$ before treatment $r_s=-.75$, $p<.0001$, latency $P_3$ after treatment $r_s=-.65$, $p<.001$, omission error rate before treatment $r_s=-.86$, $p<.0001$ and omission error rate after treatment $r_s=-.71$, $p<.0001$).

The high significant correlations were found between the latency $N_1$ before treatment and amplitude $N_1$ before treatment $r=-.61$, $p<.0001$, latency $N_1$ after treatment $r=.98$, $p<.0001$, amplitude $N_1$ after treatment $r=-.65$, $p<.0001$, latency $P_3$ before treatment $r=.75$, $p<.0001$, amplitude $P_3$ before treatment $r=-.45$, $p<.004$, latency $P_3$ after treatment $r=.67$, $p<.0001$, omission error rate before treatment $r=.72$, $p<.0001$, omission error rate after treatment $r=.46$, $p<.003$ and correlation for amplitude $P_3$ after treatment was significant at one tail $r=-.31$, $p<.052$ (for the group of ADHD children with treatment significant correlations were found between the latency $N_1$ before treatment and amplitude $N_1$ before treatment $r_s=-.79$, $p<.0001$, latency $N_1$ after treatment $r_s=.94$, $p<.0001$, amplitude $N_1$ after treatment $r_s=-.78$, $p<.0001$, latency $P_3$ before treatment $r_s=.82$, $p<.0001$, amplitude $P_3$ before treatment $r_s=-.60$, $p<.013$, latency $P_3$ after treatment $r_s=.65$, $p<.007$, amplitude $P_3$ after treatment $r_s=-.55$, $p<.027$, correlation for omission error rate after treatment was significant at one tail $r_s=.48$, $p<.059$; for the group of ADHD children without treatment significant correlations were found between the latency $N_1$ before treatment and amplitude $N_1$ before treatment $r_s=-.57$, $p<.005$, latency $N_1$ after treatment $r_s=.98$, $p<.0001$, amplitude $N_1$ after treatment $r_s=-.63$, $p<.001$, latency $P_3$ before treatment $r_s=.85$, $p<.0001$, latency $P_3$ after treatment $r_s=.74$, $p<.0001$, omission error rate after treatment $r_s=.76$, $p<.0001$, omission error rate after treatment $r_s=.61$, $p<.002$).

Table 2. Correlations of means of latency and amplitude ($N_1$ and $P_3$) before treatment for study groups.

| | | latency $N_1$ before treatment | amplitude $N_1$ before treatment | latency $N_1$ after treatment | amplitude $N_1$ after treatment | latency $P_3$ before treatment | amplitude $P_3$ before treatment | latency $P_3$ after treatment | amplitude $P_3$ after treatment | omission error rates before treatment | omission error rates after treatment |
|---|---|---|---|---|---|---|---|---|---|---|---|
| latency $N_1$ before treatment | Pearson Correlation | 1.000 | -.61(**) | .98(**) | -.65(**) | .75(**) | -.45(**) | .67(**) | -.31 | .72(**) | .46(**) |
| | Sig. (2-tailed) | | .0001 | .0001 | .0001 | .0001 | .004 | .0001 | .052 | .0001 | .003 |
| | Number of participants | | 39 | 39 | 39 | 39 | 39 | 39 | 39 | 39 | 39 |
| amplitude $N_1$ before treatment | Pearson Correlation | -.61(**) | 1.000 | -.63(**) | .93(**) | -.41(**) | .60(**) | -.45(**) | .61(**) | -.37(*) | -.47(**) |
| | Sig. (2-tailed) | .0001 | | .0001 | .0001 | .01 | .0001 | .004 | .0001 | .022 | .003 |
| | Number of participants | 39 | | 39 | 39 | 39 | 39 | 39 | 39 | 39 | 39 |
| latency $N_1$ after treatment | Pearson Correlation | .98(**) | -.63(**) | 1.000 | -.67(**) | .73(**) | -.40(*) | .66(**) | -.28 | .72(**) | .49(**) |
| | Sig. (2-tailed) | .0001 | .0001 | | .0001 | .0001 | .011 | .0001 | .08 | .0001 | .002 |
| | Number of participants | 39 | 39 | | 39 | 39 | 39 | 39 | 39 | 39 | 39 |
| amplitude $N_1$ after treatment | Pearson Correlation | -.65(**) | .93(**) | -.67(**) | 1.000 | -.45(**) | .55(**) | -.47(**) | .57(**) | -.41(**) | -.44(**) |
| | Sig. (2-tailed) | .0001 | .0001 | .0001 | | .004 | .0001 | .003 | .0001 | .01 | .005 |
| | Number of participants | 39 | 39 | 39 | | 39 | 39 | 39 | 39 | 39 | 39 |
| latency $P_3$ before | Pearson Correlation | .75(**) | -.41(**) | .73(**) | -.45(**) | 1.000 | -.43(**) | .91(**) | -.20 | .57(**) | .22 |
| treatment | Sig. (2-tailed) | .0001 | .01 | .0001 | .004 | | .007 | .0001 | .214 | .0001 | .174 |
| | Number of participants | 39 | 39 | 39 | 39 | | 39 | 39 | 39 | 39 | 39 |
| amplitude $P_3$ before treatment | Pearson Correlation | -.45(**) | .60(**) | -.40(*) | .55(**) | -.43(**) | 1.000 | -.39(*) | .77(**) | -.34(*) | -.28 |
| | Sig. (2-tailed) | .004 | .0001 | .011 | .0001 | .007 | | .015 | .0001 | .035 | .085 |
| | Number of participants | 39 | 39 | 39 | 39 | 39 | | 39 | 39 | 39 | 39 |
| latency $P_3$ after treatment | Pearson Correlation | .67(**) | -.45(**) | .66(**) | -.47(**) | .91(**) | -.39(*) | 1.000 | -.32 | .58(**) | .36(*) |
| | Sig. (2-tailed) | .0001 | .004 | .0001 | .003 | .0001 | .015 | | .051 | .0001 | .023 |
| | Number of participants | 39 | 39 | 39 | 39 | 39 | 39 | | 39 | 39 | 39 |
| amplitude $P_3$ after treatment | Pearson Correlation | -.31 | .61(**) | -.28 | .57(**) | -.20 | .77(**) | -.32 | 1.000 | -.25 | -.48(**) |
| | Sig. (2-tailed) | .052 | .0001 | .08 | .0001 | .214 | .0001 | .051 | | .129 | .002 |
| | Number of participants | 39 | 39 | 39 | 39 | 39 | 39 | 39 | | 39 | 39 |
| omission error rates before treatment | Pearson Correlation | .72(**) | -.37(*) | .72(**) | -.41(**) | .57(**) | -.34(*) | .58(**) | -.25 | 1.000 | .57(**) |
| | Sig. (2-tailed) | .0001 | .022 | .0001 | .01 | .0001 | .035 | .0001 | .129 | | .0001 |
| | Number of participants | 39 | 39 | 39 | 39 | 39 | 39 | 39 | 39 | | 39 |
| omission error rates after treatment | Pearson Correlation | .46(**) | -.47(**) | .49(**) | -.44(**) | .22 | -.28 | .36(*) | -.48(**) | .57(**) | 1.000 |
| | Sig. (2-tailed) | .003 | .003 | .002 | .005 | .174 | .085 | .023 | .002 | .0001 | |
| | Number of participants | 39 | 39 | 39 | 39 | 39 | 39 | 39 | 39 | 39 | |

The high significant correlations were found between amplitude $N_1$ before treatment and latency $N_1$ after treatment r=-.63, p<.0001, amplitude $N_1$ after treatment r=.93, p<.0001, latency $P_3$ before treatment r=-.41, p<.01, amplitude $P_3$ before treatment r=.60, p<.0001, latency $P_3$ after treatment r=-.45, p<.004, amplitude $P_3$ after treatment r=.61, p<.0001, omission error rate before treatment r=-.37, p<.022, omission error rate after treatment r=-.47, p<.003 (for the group of ADHD children with treatment significant correlations were found between amplitude $N_1$ before treatment and latency $N_1$ after treatment $r_s$=-.84, p<.0001, amplitude $N_1$ after treatment $r_s$=.97, p<.0001, latency $P_3$ before treatment $r_s$=-.83, p<.0001, amplitude $P_3$ before treatment $r_s$=.76, p<.001, latency $P_3$ after treatment $r_s$=-.69, p<.003, amplitude $P_3$ after treatment $r_s$=.70, p<.002, omission error rate before treatment $r_s$=-.50, p<.026, omission error rate after treatment $r_s$=-.77, p<.0001; for the group of ADHD children without treatment significant correlations were found between amplitude $N_1$ before treatment and latency $N_1$ after treatment $r_s$=-.60, p<.002, amplitude $N_1$ after treatment $r_s$=.88, p<.0001, latency $P_3$ before treatment $r_s$=-.57, p<.005, latency $P_3$ after treatment $r_s$=-.48, p<.022, amplitude $P_3$ after treatment $r_s$=.42, p<.045, correlation for the amplitude $P_3$ before treatment was significant at one tail $r_s$=.37, p<.08).

The high significant correlations were found between latency $N_1$ after treatment and amplitude $N_1$ after treatment r=-.67, p<.0001, latency $P_3$ before treatment r=.73, p<.0001, amplitude $P_3$ before treatment r=-.40, p<.011, latency $P_3$ after treatment r=.66, p<.0001, omission error rate before treatment r=.72, p<.0001, omission error rate after treatment r=.49, p<.002, correlation for the amplitude $P_3$ after treatment was significant at one tail

r=-.28, p<.08, (for the group of ADHD children with treatment significant correlations were found between latency $N_1$ after treatment and amplitude $N_1$ after treatment $r_s$=-80, p<.0001, latency $P_3$ before treatment $r_s$=.81, p<.0001, amplitude $P_3$ before treatment $r_s$=-.62, p<.011, latency $P_3$ after treatment $r_s$=.67, p<.004, amplitude $P_3$ after treatment $r_s$=-.55, p<.028, correlations for the omission error rate before treatment $r_s$=.44, p<.086 and omission error rate after treatment $r_s$=.47, p<.064 were significant at one tail; for the group of ADHD children without treatment significant correlations were found between latency $N_1$ after treatment and amplitude $N_1$ after treatment $r_s$=-.65, p<.001, latency $P_3$ before treatment $r_s$=.83, p<.0001, latency $P_3$ after treatment $r_s$=.70, p<.0001, omission error rate before treatment $r_s$=.76, p<.0001, omission error rate after treatment $r_s$=.64, p<.001).

The high significant correlations were found between the amplitude $N_1$ after treatment and latency $P_3$ before treatment r=-.45, p<.004, amplitude $P_3$ before treatment r=.55, p<.0001, latency $P_3$ after treatment r=-.47, p<.003, amplitude $P_3$ after treatment r=.57, p<.0001, omission error rate before treatment r=-.41, p<.01, omission error rate after treatment r=-.44, p<.005 (for the group of ADHD children with treatment significant correlations were found between the amplitude $N_1$ after treatment and latency $P_3$ before treatment $r_s$=-.77, p<.001, amplitude $P_3$ before treatment $r_s$=.64, p<.007, latency $P_3$ after treatment $r_s$=-.61, p<.012, amplitude $P_3$ after treatment $r_s$=.63, p<.01, omission error rate before treatment $r_s$=-.52, p<.038, omission error rate after treatment $r_s$=-.79, p<.0001; for the group of ADHD children without treatment significant correlations were found between the amplitude $N_1$ after treatment and latency $P_3$ before treatment $r_s$=-.61, p<.002, amplitude $P_3$ before treatment $r_s$=.44, p<.034, latency $P_3$ after treatment $r_s$=-.56, p<.005, amplitude $P_3$ after treatment $r_s$=.53, p<.009 The correlation for the omission error rate before treatment was significant at one tail $r_s$=-.37, p<.082).

Table 3. Correlations of the mean age and means of latency and amplitude ($N_1$ and $P_3$) before treatment for the group of ADHD children with treatment.

| | | latency $N_1$ before treatment | amplitude $N_1$ before treatment | latency $N_1$ after treatment | amplitude $N_1$ after treatment | latency $P_3$ before treatment | amplitude $P_3$ before treatment | latency $P_3$ after treatment | amplitude $P_3$ after treatment | omission error rates before treatment | omission error rates after treatment |
|---|---|---|---|---|---|---|---|---|---|---|---|
| latency $N_1$ before treatment | Spearman's rho | 1.000 | -.79(**) | .94(**) | -.78(**) | .82(**) | -.60(*) | .65(**) | -.55(*) | .39 | .48 |
| | Sig. (2-tailed) | | .0001 | .0001 | .0001 | .0001 | .013 | .007 | .027 | .140 | .059 |
| | Number of participants | | 16 | 16 | 16 | 16 | 16 | 16 | 16 | 16 | 16 |
| amplitude $N_1$ before treatment | Spearman's rho | -.79(**) | 1.000 | -.84(**) | .97(**) | -.83(**) | .76(**) | -.69(**) | .70(**) | -.55(*) | -.77(**) |
| | Sig. (2-tailed) | .0001 | | .0001 | .0001 | .0001 | .001 | .003 | .002 | .026 | .0001 |
| | Number of participants | 16 | | 16 | 16 | 16 | 16 | 16 | 16 | 16 | 16 |
| latency $N_1$ after treatment | Spearman's rho | .94(**) | -.84(**) | 1.000 | -.80(**) | .81(**) | -.62(*) | .67(**) | -.55(*) | .44 | .47 |
| | Sig. (2-tailed) | .0001 | .0001 | | .0001 | .0001 | .011 | .004 | .028 | .086 | .064 |
| | Number of participants | 16 | 16 | | 16 | 16 | 16 | 16 | 16 | 16 | 16 |
| amplitude $N_1$ after treatment | Spearman's rho | -.78(**) | .97(**) | -.80(**) | 1.000 | -.77(**) | .64(**) | -.61(*) | .63(**) | -.52(*) | -.79(**) |
| | Sig. (2-tailed) | .0001 | .0001 | .0001 | | .001 | .007 | .012 | .01 | .038 | .0001 |
| | Number of participants | 16 | 16 | 16 | | 16 | 16 | 16 | 16 | 16 | 16 |
| latency $P_3$ before | Spearman's rho | .82(**) | -.83(**) | .81(**) | -.77(**) | 1.000 | -.72(**) | .87(**) | -.62(**) | .61(*) | .49 |
| treatment | Sig. (2-tailed) | .0001 | .0001 | .0001 | .001 | | .002 | .0001 | .01 | .013 | .053 |
| | Number of participants | 16 | 16 | 16 | 16 | | 16 | 16 | 16 | 16 | 16 |
| amplitude $P_3$ before treatment | Spearman's rho | -.60(*) | .76(**) | -.62(*) | .64(**) | -.72(**) | 1.000 | -.66(**) | .84(**) | -.72(**) | -.60(*) |
| | Sig. (2-tailed) | .013 | .001 | .011 | .007 | .002 | | .006 | .0001 | .002 | .015 |
| | Number of participants | 16 | 16 | 16 | 16 | 16 | | 16 | 16 | 16 | 16 |
| latency $P_3$ after treatment | Spearman's rho | .65(**) | -.69(**) | .67(**) | -.61(*) | .87(**) | -.66(**) | 1.000 | -.56(*) | .50(*) | .33 |
| | Sig. (2-tailed) | .007 | .003 | .004 | .012 | .0001 | .006 | | .025 | .048 | .211 |
| | Number of participants | 16 | 16 | 16 | 16 | 16 | 16 | | 16 | 16 | 16 |
| amplitude $P_3$ after treatment | Spearman's rho | -.55(*) | .70(**) | -.55(*) | .63(**) | -.62(**) | .84(**) | -.56(*) | 1.000 | -.50(*) | -.60(*) |
| | Sig. (2-tailed) | .027 | .002 | .028 | .01 | .01 | .0001 | .025 | | .047 | .014 |
| | Number of participants | 16 | 16 | 16 | 16 | 16 | 16 | 16 | | 16 | 16 |
| omission error rates before treatment | Spearman's rho | .386 | -.55(*) | .44 | -.52(*) | .61(*) | -.72(**) | .50(*) | -.50(*) | 1.000 | .47 |
| | Sig. (2-tailed) | .140 | .026 | .086 | .038 | .013 | .002 | .048 | .047 | | .068 |
| | Number of participants | 16 | 16 | 16 | 16 | 16 | 16 | 16 | 16 | | 16 |
| omission error rates after treatment | Spearman's rho | .48 | -.77(**) | .47 | -.79(**) | .49 | -.60(*) | .33 | -.60(*) | .47 | 1.000 |
| | Sig. (2-tailed) | .059 | .0001 | .064 | .0001 | .053 | .015 | .211 | .014 | .068 | |
| | Number of participants | 16 | 16 | 16 | 16 | 16 | 16 | 16 | 16 | 16 | |

The high significant correlations were found between the latency $P_3$ before treatment and amplitude $P_3$ before treatment r=-.43, p<.007, latency $P_3$ after treatment r=.91, p<.0001, omission error rate before treatment r=.57, p<.0001 (for the group of ADHD children with treatment significant correlations were found between the latency $P_3$ before treatment and amplitude $P_3$ before treatment $r_s$=-.72, p<.002, latency $P_3$ after treatment $r_s$=.87, p<.0001, amplitude $P_3$ after treatment $r_s$=-.62, p<.01, omission error rate before treatment $r_s$=.61, p<.013. The correlation for the omission error rate after treatment was significant at one tail $r_s$=.49, p<.053; for the group of ADHD children without treatment significant correlations were found between the latency $P_3$ before treatment and amplitude $P_3$ before treatment $r_s$=-.44, p<.038, latency $P_3$ after treatment $r_s$=.90, p<.0001, omission error rate before treatment $r_s$=.59, p<.003, omission error rate after treatment $r_s$=.52, p<.01. The correlation for the amplitude $P_3$ after treatment was significant at one tail $r_s$=-.37, p<.081).

The high significant correlations were found between the amplitude $P_3$ before treatment and latency $P_3$ after treatment r=-.39, p<.015, amplitude $P_3$ after treatment r=.77, p<.0001, omission error rate before treatment r=-.34, p<.035, correlation for the omission error rate after treatment was significant at one tail r=-.28, p<.085 (for the group of ADHD children with treatment significant correlations were found between amplitude $P_3$ before treatment and latency $P_3$ after treatment $r_s$=-.66, p<.006, amplitude $P_3$ after treatment $r_s$=.84, p<.0001, omission error rate before treatment $r_s$=-.72, p<.002, omission error rate after treatment $r_s$=-.60, p<.015. The correlation for the omission error rate after treatment was significant at one tail $r_s$=.49, p<.053; for the group of ADHD children without treatment significant correlations were found between the amplitude $P_3$ before treatment and amplitude $P_3$ after treatment $r_s$=.91, p<.0001).

The high significant correlations were found between the latency $P_3$ after treatment and omission error rate before treatment r=.58, p<.0001, omission error rate after treatment r=.36, p<.023, correlation for the amplitude $P_3$ after treatment was significant at one tail r=-.32, p<.051, (for the group of ADHD children with treatment significant correlations were found between the latency $P_3$ after treatment and amplitude $P_3$ after treatment $r_s$=-.56, p<.025, omission error rate before treatment $r_s$=.50, p<.048; for the group of ADHD children without treatment significant correlations were found between the latency $P_3$ after treatment and, omission error rate before treatment $r_s$=.58, p<.004, omission error rate after treatment $r_s$=.46, p<.026, correlation for the amplitude $P_3$ after treatment was significant at one tail $r_s$=-.36, p<.093).

The high significant correlations were found between the amplitude $P_3$ after treatment and omission error rate after treatment r=-.48, p<.002 (for the group of ADHD children with treatment significant correlations were found between the amplitude $P_3$ after treatment and omission error rate after treatment $r_s$=-.50, p<.047 correlation for the amplitude $P_3$ after treatment was significant at one tail $r_s$=-.60, p<.014; for the group of ADHD children without treatment correlation analysis didn't reveal significant correlations between the amplitude $P_3$ after treatment and omission error rates).

The high significant correlations were found between the omission error rate before treatment and omission error rate after treatment r=-.57, p<.0001 (for the group of ADHD children with treatment correlations between omission error rates was significant at one tail $r_s$=.47, p<.068; for the group of ADHD children without treatment correlation for omission error rates was $r_s$=.78, p<.0001).

Table 4. Correlations of the mean age and means of latency and amplitude ($N_1$ and $P_3$) before treatment for the group of ADHD children without treatment.

| | | latency $N_1$ before treatment | amplitude $N_1$ before treatment | latency $N_1$ after treatment | amplitude $N_1$ after treatment | latency $P_3$ before treatment | amplitude $P_3$ before treatment | latency $P_3$ after treatment | amplitude $P_3$ after treatment | omission error rates before treatment | omission error rates after treatment |
|---|---|---|---|---|---|---|---|---|---|---|---|
| latency $N_1$ before treatment | Spearman's rho | 1.000 | -.79(**) | .94(**) | -.78(**) | .82(**) | -.60(*) | .65(**) | -.55(*) | .39 | .48 |
| | Sig. (2-tailed) | | .0001 | .0001 | .0001 | .0001 | .013 | .007 | .027 | .140 | .059 |
| | Number of participants | | 16 | 16 | 16 | 16 | 16 | 16 | 16 | 16 | 16 |
| amplitude $N_1$ before treatment | Spearman's rho | -.79(**) | 1.000 | -.84(**) | .97(**) | -.83(**) | .76(**) | -.69(**) | .70(**) | -.55(*) | -.77(**) |
| | Sig. (2-tailed) | .0001 | | .0001 | .0001 | .0001 | .001 | .003 | .002 | .026 | .0001 |
| | Number of participants | 16 | | 16 | 16 | 16 | 16 | 16 | 16 | 16 | 16 |
| latency $N_1$ after treatment | Spearman's rho | .94(**) | -.84(**) | 1.000 | -.80(**) | .81(**) | -.62(*) | .67(**) | -.55(*) | .44 | .47 |
| | Sig. (2-tailed) | .0001 | .0001 | | .0001 | .0001 | .011 | .004 | .028 | .086 | .064 |
| | Number of participants | 16 | 16 | | 16 | 16 | 16 | 16 | 16 | 16 | 16 |
| amplitude $N_1$ after treatment | Spearman's rho | -.78(**) | .97(**) | -.80(**) | 1.000 | -.77(**) | .64(**) | -.61(*) | .63(**) | -.52(*) | -.79(**) |
| | Sig. (2-tailed) | .0001 | .0001 | .0001 | | .001 | .007 | .012 | .01 | .038 | .0001 |
| | Number of participants | 16 | 16 | 16 | | 16 | 16 | 16 | 16 | 16 | 16 |
| latency $P_3$ before treatment | Spearman's rho | .82(**) | -.83(**) | .81(**) | -.77(**) | 1.000 | -.72(**) | .87(**) | -.62(**) | .61(*) | .49 |
| | Sig. (2-tailed) | .0001 | .0001 | .0001 | .001 | | .002 | .0001 | .01 | .013 | .053 |
|  | Number of participants | 16 | 16 | 16 | 16 |  | 16 | 16 | 16 | 16 | 16 |
| amplitude $P_3$ before treatment | Spearman's rho | -.60(*) | .76(**) | -.62(*) | .64(**) | -.72(**) | 1.000 | -.66(**) | .84(**) | -.72(**) | -.60(*) |
|  | Sig. (2-tailed) | .013 | .001 | .011 | .007 | .002 |  | .006 | .0001 | .002 | .015 |
|  | Number of participants | 16 | 16 | 16 | 16 | 16 |  | 16 | 16 | 16 | 16 |
| latency $P_3$ after treatment | Spearman's rho | .65(**) | -.69(**) | .67(**) | -.61(*) | .87(**) | -.66(**) | 1.000 | -.56(*) | .50(*) | .33 |
|  | Sig. (2-tailed) | .007 | .003 | .004 | .012 | .0001 | .006 |  | .025 | .048 | .211 |
|  | Number of participants | 16 | 16 | 16 | 16 | 16 | 16 |  | 16 | 16 | 16 |
| amplitude $P_3$ after treatment | Spearman's rho | -.55(*) | .70(**) | -.55(*) | .63(**) | -.62(**) | .84(**) | -.56(*) | 1.000 | -.50(*) | -.60(*) |
|  | Sig. (2-tailed) | .027 | .002 | .028 | .01 | .01 | .0001 | .025 |  | .047 | .014 |
|  | Number of participants | 16 | 16 | 16 | 16 | 16 | 16 | 16 |  | 16 | 16 |
| omission error rates before treatment | Spearman's rho | .386 | -.55(*) | .44 | -.52(*) | .61(*) | -.72(**) | .50(*) | -.50(*) | 1.000 | .47 |
|  | Sig. (2-tailed) | .140 | .026 | .086 | .038 | .013 | .002 | .048 | .047 |  | .068 |
|  | Number of participants | 16 | 16 | 16 | 16 | 16 | 16 | 16 | 16 |  | 16 |
| omission error rates after treatment | Spearman's rho | .48 | -.77(**) | .47 | -.79(**) | .49 | -.60(*) | .33 | -.60(*) | .47 | 1.000 |
|  | Sig. (2-tailed) | .059 | .0001 | .064 | .0001 | .053 | .015 | .211 | .014 | .068 |  |
|  | Number of participants | 16 | 16 | 16 | 16 | 16 | 16 | 16 | 16 | 16 |  |

The high significant correlations were found between the latency $P_3$ before treatment and amplitude $P_3$ before treatment r=-.43, p<.007, latency $P_3$ after treatment r=.91, p<.0001, omission error rate before treatment r=.57, p<.0001 (for the group of ADHD children with treatment significant correlations were found between the latency $P_3$ before treatment and amplitude $P_3$ before treatment $r_s$=-.72, p<.002, latency $P_3$ after treatment $r_s$=.87, p<.0001, amplitude $P_3$ after treatment $r_s$=-.62, p<.01, omission error rate before treatment $r_s$=.61, p<.013. The correlation for the omission error rate after treatment was significant at one tail $r_s$=.49, p<.053; for the group of ADHD children without treatment significant correlations were found between the latency $P_3$ before treatment and amplitude $P_3$ before treatment $r_s$=-.44, p<.038, latency $P_3$ after treatment $r_s$=.90, p<.0001, omission error rate before treatment $r_s$=.59, p<.003, omission error rate after treatment $r_s$=.52, p<.01. The correlation for the amplitude $P_3$ after treatment was significant at one tail $r_s$=-.37, p<.081).

The high significant correlations were found between the amplitude $P_3$ before treatment and latency $P_3$ after treatment r=-.39, p<.015, amplitude $P_3$ after treatment r=.77, p<.0001, omission error rate before treatment r=-.34, p<.035, correlation for the omission error rate after treatment was significant at one tail r=-.28, p<.085 (for the group of ADHD children with treatment significant correlations were found between amplitude $P_3$ before treatment and latency $P_3$ after treatment $r_s$=-.66, p<.006, amplitude $P_3$ after treatment $r_s$=.84, p<.0001, omission error rate before treatment $r_s$=-.72, p<.002, omission error rate after treatment $r_s$=-.60, p<.015. The correlation for the omission error rate after treatment was significant at one tail $r_s$=.49, p<.053; for the group of ADHD children without treatment significant correlations were found between the amplitude $P_3$ before treatment and amplitude $P_3$ after treatment $r_s$=.91, p<.0001).

The high significant correlations were found between the latency $P_3$ after treatment and omission error rate before treatment r=.58, p<.0001, omission error rate after treatment r=.36, p<.023, correlation for the amplitude $P_3$ after treatment was significant at one tail r=-.32, p<.051, (for the group of ADHD children with treatment significant correlations were found between the latency $P_3$ after treatment and amplitude $P_3$ after treatment $r_s$=-.56, p<.025, omission error rate before treatment $r_s$=.50, p<.048; for the group of ADHD children without treatment significant correlations were found between the latency $P_3$ after treatment and, omission error rate before treatment $r_s$=.58, p<.004, omission error rate after treatment $r_s$=.46, p<.026, correlation for the amplitude $P_3$ after treatment was significant at one tail $r_s$=-.36, p<.093).

The high significant correlations were found between the amplitude $P_3$ after treatment and omission error rate after treatment r=-.48, p<.002 (for the group of ADHD children with treatment significant correlations were found between the amplitude $P_3$ after treatment and omission error rate after treatment $r_s$=-.50, p<.047 correlation for the amplitude $P_3$ after treatment was significant at one tail $r_s$=-.60, p<.014; for the group of ADHD children without treatment correlation analysis didn't reveal significant correlations between the amplitude $P_3$ after treatment and omission error rates).

The high significant correlations were found between the omission error rate before treatment and omission error rate after treatment r=-.57, p<.0001 (for the group of ADHD children with treatment correlations between omission error rates was significant at one tail $r_s$=.47, p<.068; for the group of ADHD children without treatment correlation for omission error rates was $r_s$=.78, p<.0001).

Table 4. Correlations of the mean age and means of latency and amplitude ($N_1$ and $P_3$) before treatment for the group of ADHD children without treatment.

| | | latency $N_1$ before treatment | amplitude $N_1$ before treatment | latency $N_1$ after treatment | amplitude $N_1$ after treatment | latency $P_3$ before treatment | amplitude $P_3$ before treatment | latency $P_3$ after treatment | amplitude $P_3$ after treatment | omission error rates before treatment | omission error rates after treatment |
|---|---|---|---|---|---|---|---|---|---|---|---|
| latency $N_1$ before treatment | Spearman's rho | 1.000 | -.57(**) | .98(**) | -.63(**) | .85(**) | -.33 | .74(**) | -.25 | .76(**) | .61(**) |
| | Sig. (2-tailed) | | .005 | .0001 | .001 | .0001 | .122 | .0001 | .243 | .0001 | .002 |
| | Number of participants | | 23 | 23 | 23 | 23 | 23 | 23 | 23 | 23 | 23 |
| amplitude $N_1$ before treatment | Spearman's rho | -.57(**) | 1.000 | -.60(**) | .88(**) | -.57(**) | .37 | -.48(*) | .42(*) | -.29 | -.23 |
| | Sig. (2-tailed) | .005 | | .002 | .0001 | .005 | .080 | .022 | .045 | .182 | .303 |
| | Number of participants | 23 | | 23 | 23 | 23 | 23 | 23 | 23 | 23 | 23 |
| latency $N_1$ after treatment | Spearman's rho | .98(**) | -.60(**) | 1.000 | -.65(**) | .83(**) | -.30 | .70(**) | -.23 | .76(**) | .64(**) |
| | Sig. (2-tailed) | .0001 | .002 | | .001 | .0001 | .171 | .0001 | .288 | .0001 | .001 |
| | Number of participants | 23 | 23 | | 23 | 23 | 23 | 23 | 23 | 23 | 23 |
| amplitude $N_1$ after treatment | Spearman's rho | -.63(**) | .88(**) | -.65(**) | 1.000 | -.61(**) | .44(*) | -.56(**) | .53(**) | -.37 | -.27 |
| | Sig. (2-tailed) | .001 | .0001 | .001 | | .002 | .034 | .005 | .009 | .082 | .220 |
| | Number of participants | 23 | 23 | 23 | | 23 | 23 | 23 | 23 | 23 | 23 |
| latency $P_3$ before | Spearman's rho | .85(**) | -.57(**) | .83(**) | -.61(**) | 1.000 | -.44(*) | .90(**) | -.37 | .59(**) | .52(*) |

| | | | | | | | | | | |
|---|---|---|---|---|---|---|---|---|---|---|
| treatment | Sig. (2-tailed) | .0001 | .005 | .000 | .002 | | .038 | .0001 | .081 | .003 | .011 |
| | Number of participants | 23 | 23 | 23 | 23 | | 23 | 23 | 23 | 23 | 23 |
| amplitude $P_3$ before treatment | Spearman's rho | -.33 | .37 | -.30 | .44(*) | -.44(*) | 1.000 | -.35 | .91(**) | -.21 | -.23 |
| | Sig. (2-tailed) | .122 | .080 | .171 | .034 | .038 | | .107 | .0001 | .328 | .283 |
| | Number of participants | 23 | 23 | 23 | 23 | 23 | | 23 | 23 | 23 | 23 |
| latency $P_3$ after treatment | Spearman's rho | .74(**) | -.48(*) | .70(**) | -.56(**) | .90(**) | -.35 | 1.000 | -.36 | .58(**) | .46(*) |
| | Sig. (2-tailed) | .0001 | .022 | .0001 | .005 | .0001 | .107 | | .093 | .004 | .026 |
| | Number of participants | 23 | 23 | 23 | 23 | 23 | 23 | | 23 | 23 | 23 |
| amplitude $P_3$ after treatment | Spearman's rho | -.25 | .42(*) | -.23 | .53(**) | -.37 | .91(**) | -.36 | 1.000 | -.14 | -.16 |
| | Sig. (2-tailed) | .243 | .045 | .288 | .009 | .081 | .0001 | .093 | | .515 | .475 |
| | Number of participants | 23 | 23 | 23 | 23 | 23 | 23 | 23 | | 23 | 23 |
| omission error rates before treatment | Spearman's rho | .76(**) | -.29 | .76(**) | -.37 | .59(**) | -.21 | .58(**) | -.14 | 1.000 | .78(**) |
| | Sig. (2-tailed) | .0001 | .182 | .0001 | .082 | .003 | .328 | .004 | .515 | | .0001 |
| | Number of participants | 23 | 23 | 23 | 23 | 23 | 23 | 23 | 23 | | 23 |
| omission error rates after treatment | Spearman's rho | .61(**) | -.23 | .64(**) | -.27 | .52(*) | -.23 | .46(*) | -.16 | .78(**) | 1.000 |
| | Sig. (2-tailed) | .002 | .303 | .001 | .220 | .011 | .283 | .026 | .475 | .0001 | |
| | Number of participants | 23 | 23 | 23 | 23 | 23 | 23 | 233 | 23 | 23 | |

**Discussion:** As we explained above N1 reflects the simple process of recognition of stimuli. Our study replicate the position of Barry[18] and Oades [32] who confirmed that N1 latency on target stimuli showed tendency to be smaller in ADHD than controls during oddball task. According to our results it was obvious that NF does not affect on N1 parameters. Thus it means that NF serving the improvement of attention parameters can not change the auditory information recognition process in the brain. As for later response like P3 as it was mentioned in the results that the mean amplitude $P_3$ after treatment was significantly higher (p<.0001) for the ADHD treated group (mean=9.81, SD=1.87) compared to the ADHD non treated group (mean=7.3, SD=1.69). As for latency it was described that mean latency after treatment showed a significant effects of condition (p<.0001) represents a large effect size and of interaction between group and condition ( p<.0001).

Our study confirms once more the facts that in ADHD children younger than 12 years of age the amplitude of P3 on target stimuli is reduced compared to controls [33] but latency is increased [34], [35]. But the most important finding of our study was that NF training causes significant increase of amplitude and decrease of latency of P3. This point is very important as we mentioned above P3 amplitude reflects effortfulness of the stimulus response and the intensity of processing whereas P3 latency is taken as a reflection of the speed of information processing.

Thus decreasing of effortfulness of mental activity and improvement of speed of information processing is one of the most significant domain of activity in ADHD children. Modification of its component is extremely important in ADHD children as it is known that dysfunctions in cortical as well as subcortical structures plays the key role in the pathogenesis of ADHD [36]. As for psychological point of view the disorder of selection of adequate action is frequently expressed in ADHD population thus causing the impairment of executive attention. The effectiveness of NF on P3 can be explained by the following fact. It is known that NF influences neural networks that support attention, executive functions and motor regulation. According to Makris and colleagues [37] this complex network consists of several parallel networks: cortico-striatal, cortico-pallidal and cortico-cerebellar. The effectiveness of NF in the change of P3 parameters can be explained by the fact that structures functioning of which is changed during NF treatment participate in the electrogenesis of P3.

1. The role of prefrontal cortex in the pathogenesis of ADHD is evidence based [38]. It was suggested that during the occurrence of P3 the prefrontal cortex temporarily inhibits the mesencephalic activating system [39]. According to Knight the cessation of this inhibition, following the lesion of the prefrontal area could account for the P3 changes [40] detected in our study. Thus it can be marker of functioning of prefrontal areas which are extremely important in the pathogenesis of ADHD.

2. The role of parietal lobes in the pathogenesis of ADHD is confirmed by Makris and collegues [41]. According to Simson [42] although existence of a frontal P3 generator could not be excluded, the generator of the P3 could most likely be localized to the inferior parietal lobe. Knight detected that [43] upper part of the parietal lobe does not play significant role in the genesis of the auditory P3 component. On the other hand, the auditory association cortical areas at the junction of the temporo-parietal lobes seem to be very important. The anatomical connections of this area with the prefrontal cortex may be significant as it is most important part of ADHD pathogenesis.Temporal lobe and parts of limbic system are also considered as generators of P3. Halgren and colleagues [44] recorded auditory ERP from the amygdala, hippocampus and vertex using oddball paradigm. The author detecting large amplitude components from these structures suggested that they can be considered as generators of P3. The role of the hippocampus and amygdala in attention-processing has been elucidated by various clinical studies. According to Plessen and colleagues [45] connectivity of the prefrontal regions with the hippocampus and amygdala regulates a variety of attentional, memory and emotional processes implicated in the pathophysiology of ADHD. The same authors detected larger hippocampal volumes in ADHD children and reported that, among children with hippocampal dysfunction, larger volumes tended to accompany less severe ADHD symptoms. The significance of hippocampus in NF effectiveness is huge as it is proved that frontal midline theta rhythm in human EEG is often associated with hippocampal theta activity. As it is obvious that Papez circle structures like hippocampus, mammilary body of hypothalamus and anterior nucleus of thalamus all together generate theta rhythm which was the basis of attention training during NF therapy [10]. The role of temporal lobes in P3 generation was described by various authors. According to Knight [46] after unilateral temporal lobe lesion (involving area 22, area 40, and the lower part of area 39) the P3 disappear. Daruna and colleagues found that [47] in epileptic patients following unilateral temporal lobectomy the amplitude of the recorded P3 was smaller over the affected side.

3. As for basal ganglias their role in the generation of P3 was explained by Yingling [48] and Kropotov [49], [50] where striatum, globus pallidus, the ventro-lateral nucleus of the thalamus and adjacent areas were considered as a possible generators of P300 together with cortex. The fact that basal ganglia play the important role in the pathogenesis of ADHD is obvious. A growing number of neuroimaging studies have demonstrated smaller size of nucleus caudatus [51] and globus pallidus [52], [53] in ADHD children. Lubar [54] suggested that during NF therapy the basis of which is EEG activity generated by thalamus the thalamic pacemaker generates different brain rhythms depending on which cortical loops they activate but changes in cortical loops can modify the firing rate of thalamic pacemakers and hence alter their firing pattern which is the basis of NF training used in our study. Thus thalamus together with other subcortical plays the key role in ADHD.

Another important finding of our study was improvement is the rate of omission and commission errors which as we mentioned above are thought to be the index of impulsivity which together with inattentiveness is the most prominent feature of ADHD . The mean omission errors rate after treatment was significantly higher ($p<.0001$) for the ADHD non treated group (mean=4.7, SD=1.79) compared to the ADHD treated group (mean=2.38, SD=.89). The mean commission errors rate after treatment was significantly higher ($p<.0001$) for the ADHD non treated group (mean=4.87, SD=2.91) compared to the ADHD treated group (mean=.88, SD=.89). The same results were obtained for these variables using the nonparametric Mann Whitney U test. Impulsivity in ADHD children is caused by hypofunction of the dopaminergic pathways especially those projected in fronto-striatal structures.

As we have observed the high significant correlations between age and latency $N_1$ as well as between age and latency of P3 we have concluded that ERPs are in correlation with age: the smaller child has longer latency and opposite. This detection is very important as earlier start of NF treatment in children will be more effective and helps to avoid the problems accompanying ADHD in adolescence. Thus it means that NF directed to improvement of attention parameters does not affect simple auditory information processing in the human brain when significantly changes the functioning of the executive system reflected in the late ERP components connected to attention processes and selection of action which is very important part of executive functioning.

In spite of several successful findings our work need further confirmation as the follow-up study is needed to observe how consistent is improvement of P3 parameters and how long it persists. Besides it will be very interesting to identify the profile of P3 parameters in various subtypes of ADHD and level of effectiveness of NF in these subtypes of children.